\documentclass[twocolumn,secnumarabic,amssymb,superscriptaddress, nobibnotes, nofootinbib,aps,prd]{revtex4-1}

\usepackage{mathrsfs}
\usepackage{physics}
\usepackage{tikz}
\usepackage[caption=false]{subfig}
\usepackage[colorlinks=true,linkcolor=blue,citecolor=blue,urlcolor=blue]{hyperref}
\usepackage{cleveref}
\usepackage{amsmath,amssymb,amsfonts,amsthm}
\usepackage{graphicx}
\usepackage{ragged2e}
\usepackage{float}
\usepackage{bm}
\usepackage{svg}

\usepackage{dsfont}
\usepackage[title]{appendix}

\usepackage{hyperref}

\def\bea{\begin{eqnarray}}
\def\eea{\end{eqnarray}}
\def\be{\begin{equation}}
\def\ee{\end{equation}}

\usepackage{filecontents}

\begin{document}

\title{Reliable quantum master equation of the Unruh-DeWitt detector}%

\author{Si-Wei Han}
\affiliation{School of Physics, Xi'an Jiaotong University, Xi'an, Shaanxi 710049, China}

\author{Wenjing Chen}
\affiliation{School of Physics, Xi'an Jiaotong University, Xi'an, Shaanxi 710049, China}

\author{Langxuan Chen}
\affiliation{School of Physics, Xi'an Jiaotong University, Xi'an, Shaanxi 710049, China}

\author{Zhichun Ouyang}
\affiliation{Department of Physics, Hong Kong University of Science and Technology, \\
Clear Water Bay, Hong Kong, China}

\author{Jun Feng}%
\email{Corresponding author: j.feng@xjtu.edu.cn}
\affiliation{School of Physics, Xi'an Jiaotong University, Xi'an, Shaanxi 710049, China}
\affiliation{Hefei National Laboratory, Hefei 230088, Anhui, China}

\date{\today}%

\begin{abstract}
In this paper, we present a method for estimating the validity range of the quantum Markovian master equation as applied to the Unruh-DeWitt (UDW) detector within a broader context, particularly without necessitating an exact solution for the detector's evolution. We propose a relaxed van Hove limit (i.e., late-time limit) and offer a perturbative estimate of the error order resulting from the standard derivation procedure of open quantum dynamics. Our primary findings include reliability criteria for the Markov approximation and conditions for the applicability of the rotating wave approximation. Nevertheless, the specific forms of these validity conditions rely on the details of the detector-field system, such as the spacetime background, the trajectory of the detector, and the type of quantum field being analyzed. Finally, we illustrate our results by reexamining the open dynamics of an accelerating UDW detector undergoing the Unruh effect, where the validity conditions narrow the parameter space to ensure the solution's reliability regarding the quantum Markovian master equation. 
\end{abstract}

\maketitle

\section{Introduction}
\label{1}

The particle content of a quantum field is an observer-dependent notion. For example, even in the Poincar\'e-invariant Minkowski vacuum, an observer with constant acceleration perceives a thermal spectrum \cite{sec1-1}. This celebrated "Unruh effect" can best be manifested by some particlelike quantum probes, among which the best-known idealized model is the Unruh-DeWitt (UDW) detector \cite{sec1-2,sec1-3}. Formally, it consists of a two-level particle that couples to the background quantum field in a way similar to light-matter interactions while neglecting angular momentum exchange. In a weak coupling limit, i.e., assuming the coupling constant is $g\ll1$, the response to the detector-field interaction $H_{\text{int}}$ can be solved \textit{perturbatively}. For an accelerating detector in Minkowski spacetime, its \textit{asymptotic} transition rate from the ground state to the excited state exhibits a thermal Planckian spectrum with a temperature proportional to its acceleration. This is regarded as a result of detailed balance between the absorption and emission of field quanta, thus manifesting the particle content of the background quantum field \cite{sec1-4-1,sec1-4-2}. Much work has been carried out to examine the response and thermal behavior of a detector in various contexts \cite{sec1-5}, such as black holes \cite{sec1-6,sec1-7}, expanding universe \cite{sec1-8,sec1-9}, quantum gravity \cite{sec1-10}, etc.

The drawback of above detector-field approach is that its temporal validity is restricted as the perturbation calculation works only for the range $0<t\ll{1}/{g^2}$. This is because that direct perturbation calculation via $e^{-i g H_{\text{int}} t} \simeq 1-i g H_{\mathrm{int}} t+\cdots$ involves combinations of $g H_{\text{int}} t$. Therefore, at very late time when $g H_{\text {int}} t$ becomes too large, high-order contributions of amplitudes cannot be ignored, leading to the failure of perturbation expansion. In general, any late-time prediction of UDW detector dynamics fails at $g^2 t \sim \mathcal{O}(1)$ \cite{sec1-11}, hindering our understanding of long-time processes such as thermalization, but stopping at the detailed balance condition or Planckian transition rates.
 
An alternative perspective is to treat the UDW detector within the framework of the theory of open quantum systems \cite{sec1-12}. The time evolution of the detector as a local open system is then governed by a \textit{quantum Markovian master equation} (QMME), where its interaction with background quantum fluctuations acts as the "bath" inducing dissipation and decoherence terms. In his seminal works \cite{sec1-13,sec1-14}, Davies presented a rigorous mathematical proof demonstrating that, once the bath meets general conditions, the QMME for open dynamics becomes accurate in the van Hove limit [i.e., $g^2 t \sim \mathcal{O}(1)$ while taking $g \rightarrow 0$ and $t \rightarrow \infty$ simultaneously], as all higher-order terms tend toward zero. Consequently, the open system approach is particularly powerful for tracking long-time processes, such as thermalization. In recent years, intensive works \cite{sec1-17,sec1-18,sec1-19,sec1-20,sec1-21,sec1-22,sec1-23,sec1-24,sec1-24+1} have been undertaken by applying this method to the UDW detector in various spacetimes, whose complete dynamics encode the thermal nature of Hawking-Unruh-type effects. For multi-UDW detectors, this method highlights the possibility for generating nonclassical correlations, which stem from collective open dynamics and can be observed through various correlation witnesses \cite{sec1-25,sec1-26,sec1-27,sec1-28,sec1-29,sec1-30}, such as entanglement monotones, nonlocality, and quantum uncertainty bounds, depending on detector trajectories, interdistance between detectors, the types of fields in different spacetimes, etc.

Despite its productive applications, the relevance of the master equation regarding the intricate interactions between UDW detectors and quantum fields is frequently overlooked. In fact, most studies consider the QMME a ``black box" and accept it as valid without further justification. This oversight partly arises from the challenge of assessing the validity of the QMME by directly comparing its solution to the exact one. Specifically, the precise open dynamics are shaped by a differential-integral equation, whose solution at a specific moment relies on its entire evolutionary history and is, therefore, formidable. Conversely, the QMME can establish a quantum dynamical map (e.g., in the form of the Gorini-Kossakowski-Sudarshan-Lindblad (GKSL) master equation \cite{sec1-15,sec1-16}) that connects open system states at arbitrary times, just by suitably disregarding the aforementioned memory effect through several approximations:

$(i)$~\textit{Born approximation.} In cases of weak coupling between the detector and quantum fields, it can be approximated that the state of the field remains unchanged throughout its interaction with the detector. This assumption serves to simplify the structure of the exact evolution equation; however, it remains challenging to solve due to its dependence on the entire history of the detector's evolution. 

$(ii)$~\textit{Markov approximation.} Assuming that the correlation function of the quantum field $\mathcal{W}(t)$ decays with a characteristic timescale $t_B$, e.g., $\mathcal{W}(t) \sim e^{-t / t_B}$, once the typical evolution timescale of the detector is much larger than the correlation time (i.e., $t \gg t_B$), the effective integration region over time falls within $t \in\left[0, t_B\right)$. Consequently, the detector's evolution at any specific time relies solely on its state at that moment, indicating that the memory effect of the equation has been neglected.

$(iii)$~\textit{Rotating wave approximation (RWA).} The result of the ``Born-Markov" approximation is a non-\textit{CP} (completely positive) evolution equation known as the \textit{Redfield equation}, indicating that the density matrix is notoriously mapped to nonpositive matrix. This dilemma underscores the necessity of further utilizing the RWA; all rapidly oscillating terms relative to the typical evolution timescale in the Redfield equation have a negligible effect on detector evolution and can, therefore, be disregarded. While the RWA further limits the domain of the master equation, the combined ``Born-Markov-RWA" approximation consistently ensures that the final derived QMME remains completely positive definite \cite{sec1-12}.

The validity range of QMME is constrained by the errors introduced through the use of the aforementioned approximations, which have ignited an ongoing debate about their applicability and necessity \cite{sec1-31,sec1-32,sec1-33+}. For example, a recent series of works \cite{sec1-33,sec1-34,sec1-35,sec1-36} revisited the open system framework for the UDW detector system from an open EFT perspective, which naturally introduces a hierarchy of timescales of the environment and the system, limiting each approximation to a specific domain of validity. In \cite{sec1-37}, the authors examined the leading-order expansion of the \textit{memory kernel}, which provides validity relations that explicitly bound the range of parameter space that Markov approximation allows. Furthermore, it was argued that with a carefully taken Markov approximation, the evolution map can be \textit{CP}-preserving without additional reliance on the RWA approximation. 
 
In this paper, we reassess the validity of the quantum Markovian master equation within a broader context, specifically without requiring the precise solution of the detector. We invoke the Born approximation to exclude the backreaction of the background quantum field, which is reasonable since it possesses infinite degrees of freedom spreading throughout spacetime. Instead, we address and estimate the primary error introduced by neglecting the memory effect via the Markov approximation. Assuming a small nonvanishing $g$ for the UDW detector model, we propose a relaxed van Hove limit [we refer to \textit{late-time limit} $g^2 t \sim \mathcal{O}(1)$]. By expanding the detector state in terms of its evolution history, which corresponds to a perturbative series of powers of $g$, we find that the first-order memory effect results in an error of order $g^4$. Consequently, the $g$-dependent reliability conditions [i.e., Eqs.(\ref{eq3.6})] for the Markov approximation can be established by requiring that the first-order memory effect is negligible. Under the late-time limit, we can further identify from the reliability conditions the timescale at which the master equation becomes reliable for the UDW detector-field interaction. On the other hand, unlike previous studies \cite{sec1-33,sec1-34,sec1-35,sec1-36,sec1-37}, after closely revisiting Davies' theorems on the rigorous derivation of the QMME, we show that the RWA remains necessary for restoring the \textit{CP} property of the UDW detector evolution. However, some applicability conditions [i.e., Eqs.(\ref{bujia1.3})] should primarily be satisfied.

The structure of this paper is as follows: In Sec. \ref{2}, we review the Unruh-DeWitt detector model and the derivation of its QMME. In Sec. \ref{3}, we determine the valid range of the QMME, which is summarized to two sets of conditions. Finally, in Sec. \ref{4}, we reexamine the Unruh effect as an application of our results and determine the range of physical parameters to guarantee the validity of the QMME prediction. For simplicity, throughout the paper, we use natural units $\hbar=c=k_B=1$.

\section{The dynamic equation for the Unruh-DeWitt detector}
\label{2}
We formulate the open dynamic equation of the UDW detector and examine how the QMME arises from three approximations \cite{sec2-1}; the Born, Markov, and RWA approximations.

\subsection{The setup}
\label{hsw}
A UDW detector functions as a two-level quantum system interacting with a quantum field in fixed spacetime. The detector evolves on a classical trajectory due to this interaction. The Hamiltonian of the detector-field system takes the form,
\begin{equation}
H=H_d+H_\phi+g H_{\text{int}},
\label{eq2.1}
\end{equation}
where the detector Hamiltonian $H_d=\frac{1}{2} \omega \sigma_3$ models a two-level atom with $\omega$ as its energy gap. The Hamiltonian $H_{\phi}$ of the quantum field can be determined once we specify the background spacetime geometry and the type of field being studied. The interaction Hamiltonian $H_{\text{int}}$ describes the coupling between the detector and the field, taking the form $\mathfrak{m}(0) \otimes \phi[x(t)]$, where $\mathfrak{m}$ is the monopole momentum operator of the detector, generally represented as $\left(\sigma_{+}+\sigma_{-}\right)$, and $x(t)$ denotes the classical trajectory of the detector. Furthermore, the interaction Hamiltonian is scaled by a coupling constant $g$ to control the strength of the interaction.

Working within the interaction picture, the momentum and field operators are transformed, like,
\be
\mathfrak{M}(t)=\mathcal{U}_d^{\dagger} \mathfrak{m}(0) \mathcal{U}_d~~~~, ~~~~ {\Phi}(t)=\mathcal{U}_\phi^{\dagger} \phi[x(t)] \mathcal{U}_\phi, 
\label{eq2.3}
\ee
with $\mathcal{U}_\alpha=\exp\left(-i H_d t\right)$ and $\mathcal{U}_\phi=\exp\left(-i H_\phi t\right)$, respectively. Then, the interaction Hamiltonian becomes
\begin{equation}
\tilde{H}_{\text{int}}(t)=g\;\mathfrak{M}(t) \otimes {\Phi}(t).
\label{eq2.2}
\end{equation}

Let $\rho(t)$ represent the total density matrix of the detector-field system. The evolution of $\rho(t)$ is governed by the Liouville-von Neumann equation,
\begin{equation}
\frac{d \rho(t)}{d t}=-i\left[g \tilde{H}_{\text{int}}(t), \rho(t)\right],
\label{eq2.4}
\end{equation}
which can be formally integrated into
\begin{equation}
\rho(t)=\rho(0)-i \int_0^t d s\left[g \tilde{H}_{\text{int}}(s), \rho(s)\right].
\label{eq2.5}
\end{equation}
Since only the time evolution of the UDW detector is relevant, we can trace over all field degrees of freedom from the total state (\ref{eq2.5}) to derive the dynamic equation for the detector density matrix $\rho_d(t)=\operatorname{Tr}_{\Phi}(\rho)$ as
\begin{equation}
\begin{aligned}
\dot{\rho}_d(t)&=-i \operatorname{Tr}_{\Phi}\left\{\left[g \tilde{H}_{\text{int}}(t), \rho(0)\right]\right\}\\
&- \operatorname{Tr}_{\Phi}\left\{\left[g \tilde{H}_{\text{int}}(t), \int_0^t d s\left[g \tilde{H}_{\text{int}}(s), \rho(s)\right]\right\}\right..
\end{aligned}
\label{eq2.6}
\end{equation}
Changing variables to $\tau=t-s$, so that $\int_0^t d s=-\int_t^0(-d \tau)=\int_0^t d \tau$, Eq. (\ref{eq2.6}) can be rewritten as
\begin{equation}
\begin{aligned}
\dot{\rho}_d(t)&=-i \operatorname{Tr}_{\Phi}\left\{\left[g \tilde{H}_{\text{int}}(t), \rho(0)\right]\right\}\\
&- \operatorname{Tr}_{\Phi}\left\{\left[g \tilde{H}_{\text{int}}(t), \int_0^t d \tau\left[g \tilde{H}_{\text{int}}(t-\tau), \rho(t-\tau)\right]\right\}\right..
\end{aligned}
\label{eq2.7}
\end{equation}
This equation describes the exact evolution of the detector, but it is very challenging to solve because of the integral term in the second part. To evaluate the detector state $\rho_d$ at a specific time $t$, we need to know the complete evolution history of $\rho(\tau)$ from $0$ to $t$ in the integral of Eq. (\ref{eq2.7}), which reflects the \textit{memory effect}. However, by introducing several approximations, Eq. (\ref{eq2.7}) can be reformulated into the QMME, where the memory contribution can be effectively eliminated.

\subsection{Derivation of quantum Markovian master equation}
In a ``bath" (quantum field) with significantly larger degrees than the detector, it is reasonable to assume that the detector evolves while the field remains largely unaffected. Thus, the state of the detector-field system at time $t$ takes on a product form as
\begin{equation}
\rho(t) \approx \rho_d(t) \otimes \rho_\Phi,
\label{eq2.8}
\end{equation}
where $\rho_\Phi$ is the time-independent, stationary field state. 

Under this Born approximation, Eq. (\ref{eq2.7}) becomes
\begin{widetext}
\begin{equation}
\dot{\rho}_d(t)
\approx-i \operatorname{Tr}_\Phi\left\{\left[g \tilde{H}_{\text{int}}(t), \rho_d(t) \otimes \rho_\Phi\right]\right\}
- \operatorname{Tr}_\Phi\left\{\left[g \tilde{H}_{\text {int }}(t), \int_0^t d \tau\left[g \tilde{H}_{\text {int }}(t-\tau), \rho_d(t-\tau) \otimes \rho_\Phi\right]\right]\right\}.
\label{eq2.9}
\end{equation}
The first term on the right-hand side can be reduced to
\begin{equation}
\begin{aligned}
 -i \operatorname{Tr}_\Phi\left\{\left[g \tilde{H}_{\text {int }}(t), \rho_d(t) \otimes \rho_\Phi\right]\right\} 
=&  -ig \operatorname{Tr}_\Phi\left\{ \mathfrak{M}(t) \rho_d(t) \otimes \Phi(t) \rho_\Phi\right\}+ig \operatorname{Tr}_\Phi\left\{ \rho_d(t) \mathfrak{M}(t) \otimes \rho_\Phi \Phi(t)\right\} \\
= & -i g\operatorname{Tr}_\Phi\left\{\rho_\Phi \Phi(t)\right\}\left[\mathfrak{M}(t), \rho_d(t)\right]\\
=&-ig\langle\Phi(t)\rangle\left[\mathfrak{M}(t), \rho_d(t)\right].
\end{aligned}
\label{eq2.10}
\end{equation}
For the quantum field in the vacuum state, the one-point function $\langle\Phi(t)\rangle=0$, which means that the first term of Eq.(\ref{eq2.9}) is always zero.
\end{widetext}
Expanding the second term of Eq.(\ref{eq2.9}) and after some calculations, one finally obtains
\begin{equation}
\dot{\rho}_d(t)=g^2 \int_0^t d \tau\left\{\mathcal{W}(\tau)\left[\mathfrak{M}(t-\tau) \rho_d(t-\tau),\mathfrak{M}(t)\right]+\text {H.c.}\right\}.
\label{eq2.11}
\end{equation}
Here $\mathcal{W}(\tau)$ denotes the Wightman function of the quantum field as
\begin{equation}
\mathcal{W}\left(\tau_1-\tau_2\right):=\operatorname{Tr}_\Phi\left\{\rho_\Phi \Phi\left[x\left(\tau_1\right)\right] \Phi\left[x\left(\tau_2\right)\right]\right\},
\label{eq2.12}
\end{equation}
which describes the correlation of the quantum field at various times along the detector trajectory.

We aim to derive a time-local (i.e., memoryless) differential equation for $\rho_d$, which depends solely on $t$ and not on the detector's history. To accomplish this, we assume the background field has a short correlation time $\tau_B$, meaning the corresponding Wightman function decays rapidly, such as e.g., $\left|\mathcal{W}(\tau)\right| \sim e^{-\tau / \tau_B}$, and becomes zero for $\tau \gg \tau_B$. Under this Markov approximation, the main contribution to the integral in Eq.(\ref{eq2.11}) comes from the vicinity of $\tau \approx 0$. The short ``memory" of the bath correlation function allows us to replace $\rho_d(t-\tau)\approx\rho_d(t)$ without introducing significant errors, which results in the \textit{Redfield equation},
\begin{equation}
\dot{\rho}_d=g^2 \int_0^\infty d \tau\left\{\mathcal{W}(\tau)\left[\mathfrak{M}(t-\tau) \rho_d(t),\mathfrak{M}(t)\right]+\text{H . c .}\right\}.
\label{eq2.14}
\end{equation}
Obviously, the Markov approximation could become increasingly accurate for longer relaxation timescales of the detector, says $t \gg \tau_B$.

Unfortunately, the Redfield equation is unreliable as it fails to maintain the \textit{CP} (completely positive) property of quantum dynamics, risking a nonpositive density matrix \cite{sec2-2}. One may argue that the problem stems from an improperly performed Markovian limit. Recently, it was proposed \cite{sec1-33,sec1-37} that an alternative Markov approximation, approximating both system state $\rho_d(t-\tau)$ and observables $\mathfrak{M}(t-\tau)$ as memoryless. Essentially, their alternative approach can be understood as a synergistic combination of the van Hove limit and the singular coupling limit \cite{plus-alicki}, where the latter effectively replaces $\mathcal{W}(\tau)$ in Eq.\eqref{eq2.14} with a delta-function like $\mathcal{W}(\tau) \approx \Lambda \delta(\tau)$, a treatment primarily applied in the high-temperature regime. This leads us to a different master equation, formally as
\be
\begin{aligned}
\dot{\rho}_d(t)&=g^2 \int_0^{\infty} d \tau\left\{\mathcal{W}(\tau)\left[\mathfrak{M}(t) \rho_d(t), \mathfrak{M}(t)\right]+\text {H.c. }\right\}\\
&=\frac{g^2}{2}\left\{\Lambda\left[\mathfrak{M}(t) \rho_d(t), \mathfrak{M}(t)\right]+\text {H.c. }\right\}
\end{aligned}
\ee
Under proper reliability conditions, the alternative master equation generates approximately \textit{CP} dynamics, making further approximations unnecessary \cite{sec2-1,secadd}. The alternative master equation is particularly suitable when the coupling constant is sufficiently small and the background field temperature is high enough, as justified by its reliability conditions \cite{sec1-37}. In this paper, however, we remain to utilize the standard Markov approximation without restrictions in high-temperature scenarios. To restore the \text{CP} of quantum dynamics, the third step, RWA, is necessary.

To attain this, one first converts the monopole momentum operator to the frequency domain,
\begin{equation}
\mathfrak{M}(t)=\sigma_{+} e^{i \omega t}+\sigma_{-} e^{-i \omega t}=\sum_{\alpha=\omega,-w} \mathfrak{M}_\alpha e^{-i \alpha t},
\label{eq2.15}
\end{equation}
where $\mathfrak{M}_\omega:=\sigma_{+}$ and $\mathfrak{M}_{-\omega}:=\sigma_{-}$. The two terms in the commutator $\left[\mathfrak{M}(t), \mathfrak{M}(t-\tau) \rho_d(t)\right]$ become
\begin{equation}
\begin{aligned}
 \mathfrak{M}(t) \mathfrak{M}(t-\tau) \rho_d(t)
 &=\sum_{\alpha, \alpha^{\prime}} e^{i \alpha \tau} e^{i\left(\alpha^{\prime}-\alpha\right) t} \mathfrak{M}_{\alpha^{\prime}}^{+} \mathfrak{M}_\alpha \rho_d(t), \\
\mathfrak{M}(t-\tau) \rho_d(t) \mathfrak{M}(t)
&=\sum_{\alpha, \alpha^{\prime}} e^{i \alpha \tau} e^{i\left(\alpha^{\prime}-\alpha\right) t} \mathfrak{M}_\alpha \rho_d(t) \mathfrak{M}_{\alpha^{\prime}}^{+}.
\end{aligned}
\label{eq2.17}
\end{equation}

Plugging these decompositions into the Redfield equation (\ref{eq2.14}), we obtain
\begin{equation}
\dot{\rho}_d=g^2 \sum_{\alpha, \alpha^{\prime}}\left\{\Gamma(\alpha) e^{i\left(\alpha^{\prime}-\alpha\right) t}\left[\mathfrak{M}_\alpha \rho_d(t),\mathfrak{M}_{\alpha^{\prime}}^{\dagger}\right]+\text{H . c.}\right\}.
\label{eq2.19}
\end{equation}
Here, all the $\tau$-dependent terms in Eq.(\ref{eq2.14}) have been integrated into a single function $\Gamma(\alpha)$, which is the one-sided Fourier transform of the Wightman function
\begin{equation}
\Gamma(\alpha):=\int_0^{\infty} d \tau e^{i \alpha \tau} \mathcal{W}(\tau).
\label{eq2.18}
\end{equation}
For later convenience, it is important to note \cite{sec1-13,sec1-14} that the function $\Gamma(\alpha)$ can be separated into the real and imaginary parts,
\begin{equation}
\Gamma(\alpha)=\frac{1}{2} \gamma(\alpha)+i S(\alpha),
\label{eq2.20}
\end{equation}
where $\gamma(\alpha)$ is the Fourier transform of bath correlator,
\begin{equation}
\gamma(\alpha):=\int_{-\infty}^{\infty} d \tau e^{i \alpha \tau} \mathcal{W}(\tau),
\label{eq2.24}
\end{equation}
and real function $S(\alpha)$ is defined as
\begin{equation}
S(\alpha):=\frac{1}{2 \pi} \int_{-\infty}^{\infty} \gamma\left(\alpha^{\prime}\right) \mathcal{P}\left(\frac{1}{\alpha-\alpha^{\prime}}\right) d \alpha^{\prime}=S^*(\alpha),
\end{equation}
including the Cauchy principal value $\mathcal{P}$. 

In a integral form, the Eq.(\ref{eq2.19}) can be written as
\begin{widetext}
\begin{equation}
\rho_d(t)=-g^2 \int_0^t d \tau\left(\sum_\alpha \Gamma(\alpha)\left[m_{\alpha}^{+}, m_\alpha \rho_d(\tau)\right]+\sum_{\alpha \neq \alpha^{\prime}} \Gamma(\alpha)e^{i\left(\alpha^{\prime}-\alpha\right) t}\left[m_{\alpha^{\prime}}^{+}, m_\alpha \rho_\alpha(\tau)\right]+\text{H.c.}\right),
\label{eq2.25}
\end{equation}
\end{widetext}
where the summation has been split into two parts. One can observe that when $t \gg \left|\alpha^{\prime}-\alpha\right|^{-1}$, the terms with summation over $\alpha \neq \alpha^{\prime}$ oscillate so rapidly that they average to zero. This captures the essence of RWA. Recall that the Markov master equation is valid for $t \gg \tau_B$. Likewise, as we examine a longer timescale, the oscillating terms in Eq. (\ref{eq2.25}) will appear increasingly ``fast-oscillating", which makes the RWA more accurate.

Eventually, by introducing Eq.(\ref{eq2.20}) and the RWA into the Redfield equation, we arrive at the so-called Gorini-Kossakowski-Sudarshan-Lindblad (GKSL) form of QMME as \cite{sec1-15,sec1-16}
\begin{widetext}
\begin{equation}
\frac{d \rho_d(t)}{d t}=-i\left[H_{L S}, \rho_d(t)\right]+g^2 \sum_\alpha \gamma(\alpha)\left(\mathfrak{M}_\alpha \rho_d(t) \mathfrak{M}_\alpha^{\dagger}-\frac{1}{2}\left\{\mathfrak{M}_\alpha^{\dagger} \mathfrak{M}_\alpha, \rho_d(t)\right\}\right),
\label{eq2.21}
\end{equation}
\end{widetext}
where the Lamb shift Hamiltonian is given by
\begin{equation}
H_{L S}:=g^2 \sum_\alpha S(\alpha) \mathfrak{M}_\alpha^{\dagger} \mathfrak{M}_\alpha.
\label{eq2.22}
\end{equation}
Using Bochner's theorem, one can prove \cite{sec1-12} that RWA guarantees the condition
\begin{equation}
\gamma(\alpha)>0,
\label{eq2.23}
\end{equation}
which is the sufficient and necessary condition ensuring the QMME (\ref{eq2.21}) always preserves the \textit{CP} property of detector evolution.

\section{Reliable quantum master equation}
\label{3}

The derivation of the QMME indicates that as time $t$ approaches infinity (certainly fulfills $t \gg\left|\alpha^{\prime}-\alpha\right|^{-1}$ and $t\gg\tau_B$), the errors stemming from the Markov and RWA approximations tend to vanish. However, this mathematical limit is unsatisfactory because, from a physical perspective, it is vague to refer to a ``sufficiently long” timescale or ``oscillate rapidly". This motivates us to identify a more precise timescale at which the QMME description of open dynamics \textit{becomes} physically meaningful and reliable. In this section, we analyze the errors that accumulate over time due to the Markov and RWA approximations. By the proposed relaxed van Hove limit, we establish the criteria for the reliability of the Markov approximation and for the applicability of the RWA, which are upheld in a narrower parameter space for the UDW detector.

\subsection{General consideration}

\begin{figure}[htbp]  
    \centering 
    \includegraphics[width=.4\textwidth]{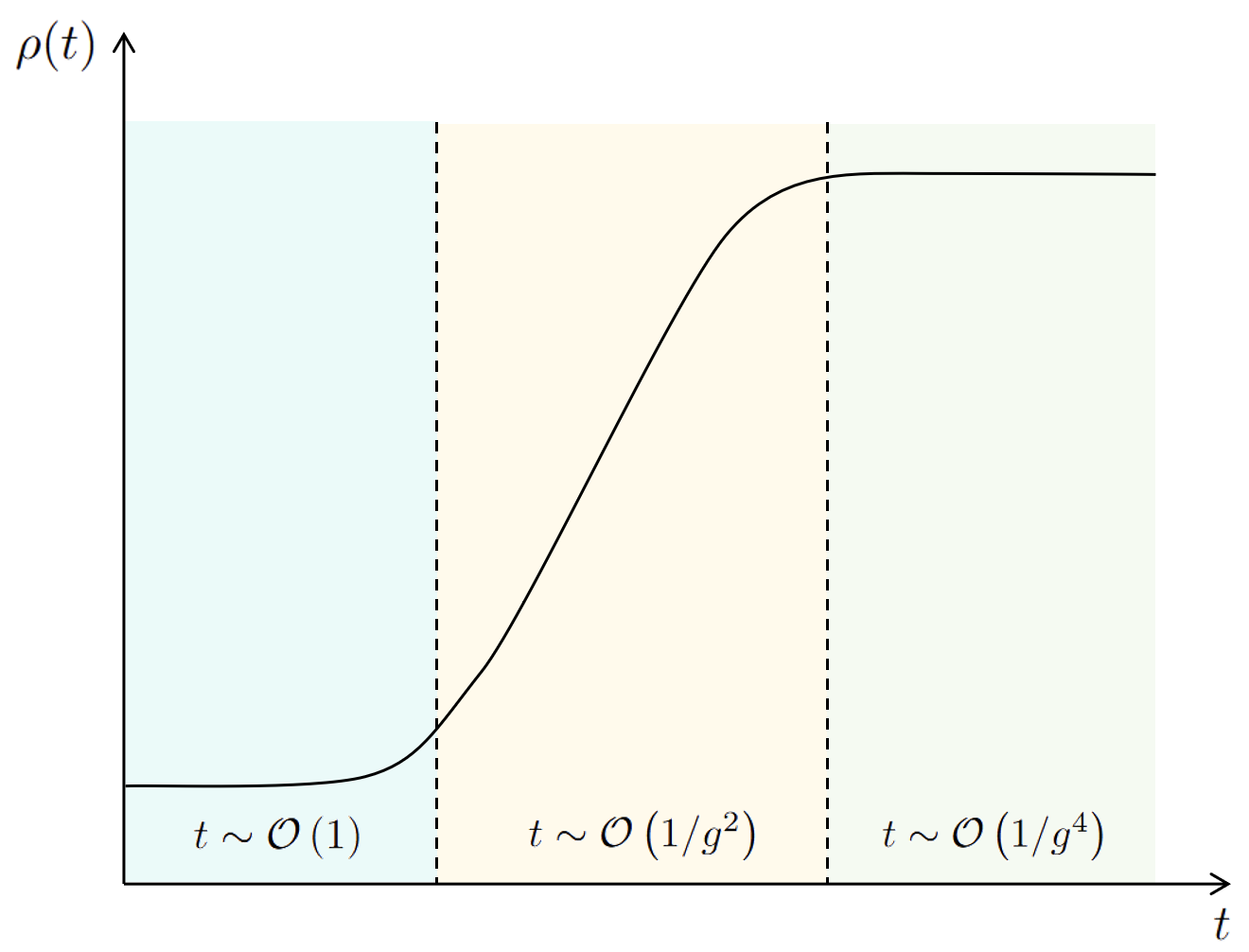}
    \caption{Schematic diagram of the evolution of $\rho_d(t)$ governed by Eq.(\ref{eq2.7}) under weak coupling. By taking the weak coupling limit $g \rightarrow 0$, the evolution of the detector is only salient once $t$ approaches $\mathcal{O}\left(1 / g^2\right)$, which suggests a genuine limit  $g^2 t \sim \mathcal{O}(1)$ while simultaneously taking $g \rightarrow 0$ and $t \rightarrow \infty$, should be taken to guarantee the convergence of Eq.(\ref{eq2.7}) \cite{sec1-13,sec1-14}. }
        \label{fig6}
\end{figure}

In his seminal works 
\cite{sec1-13,sec1-14}, Davies aimed to provide a mathematically rigorous definition of the quantum Markovian master equation within a general context. He noted that the order of Eq.(\ref{eq2.7}) is $g^2$, indicating that by taking the weak coupling limit $g \rightarrow 0$, the evolution of the open system only becomes drastic once $t$ approaches $\mathcal{O}\left(1 / g^2\right)$. For $t$ earlier than this timescale, however, the perturbative evolution is moderate under the weak coupling limit, as illustrated in Fig.\ref{fig6}. Consequently, Davies proposed a genuine limit (later known as the \textit{van Hove limit} \cite{sec3-1}), $g^2 t \sim \mathcal{O}(1)$ while simultaneously taking $g\rightarrow0$ and $t \rightarrow \infty$, under which Eq.(\ref{eq2.7}) converges to the QMME (\ref{eq2.21}) if the correlation function of the quantum field is integrable.\footnote{Strictly, this means that $\int_0^{\infty}|\mathcal{W}(\tau)|(1+\tau)^{\varepsilon} d \tau<\infty$, for certain $\varepsilon>0$, which is a rigorous manifestation of the previously mentioned requirement for the Wightman function to "decay sufficiently fast".}

We note that Davies' result is a limit theorem asserting that for a sufficiently small coupling constant, the master equation provides a good approximation of the real dynamics. However, from a physical perspective, it is unsatisfactory, as Davies' theorem lacks refined conditions to quantify what is meant by a given physical coupling being ``small enough" \cite{plus-alicki} or, equivalently, a timescale being ``sufficiently long" whenever the prediction of QMME is reliable.

In the following, we demonstrate that these conditions on the valid range of QMME can be specified by estimating the leading-order error terms of (\ref{eq2.21}). In this context, our primary interest lies in a relaxed van Hove limit, which we refer to \textit{late-time limit}, i.e., $g^2 t \sim \mathcal{O}(1)$ with small but nonvanishing $g$. We calculate higher-order error terms of the quantum master equation inherited from Eq.(\ref{eq2.7}) and require them to be much smaller than the terms in the master equation Eq.(\ref{eq2.21}). We anticipate that the derived reliability conditions should be $g$-dependent and automatically hold as $g \rightarrow 0$, as the van Hove limit has been recovered.

\subsection{Reliability conditions for the Markov approximation}

To estimate the error introduced by the Markov approximation, we compare Eq.(\ref{eq2.11}), the dynamical equation under the Born approximation, with the Redfield equation Eq.(\ref{eq2.14}) derived after applying the Markov approximation. The error resulting from the Markov approximation can be isolated and reformulated into
\begin{widetext}
\begin{equation}
\begin{aligned}
 \frac{d \rho_d(t)}{d t}&=-g^2 \int_0^t d \tau\left\{\mathcal{W}(\tau)\left[\mathfrak{M}(t), \mathfrak{M}(t-\tau) \rho_d(t-\tau)\right]+\text{H. c.}\right\} \\
&=\underbrace{-g^2 \int_0^{\infty} d \tau\left\{\mathcal{W}(\tau)\left[\mathfrak{M}(t), \mathfrak{M}(t-\tau) \rho_d(t)\right]+\text{H. c.}\right\}}_{\text {Markov }} \\
&\underbrace{+g^2\left\{ \int_0^{\infty} d\tau \mathcal{W}(\tau)\left[\mathfrak{M}(t), \mathfrak{M}(t-\tau)\left(\rho_d(t)-\rho_d(t-\tau)\right)\right]+ \int_t^{\infty} d \tau \mathcal{W}(\tau)\left[\mathfrak{M}(t), \mathfrak{M}(t-\tau) \rho_d(t-\tau)\right]\right\}+\text{H. c.}}_{\text {error part}}.
\end{aligned}
\label{eq3.1}
\end{equation}
\end{widetext}
The estimated upper bound of the second term of the error part (see Appendix \ref{A}) indicates that its contribution gradually diminishes beyond the late-time limit. This suggests that the primary source of error from neglecting the memory effect (Markov approximation) should be dominated by the first error term. 

Along the evolution history, we expand $\rho_d(t-\tau)$ as
\begin{equation}
\rho_d(t-\tau)=\rho_d(t)-\frac{d \rho_d(t)}{d t} \tau+\frac{1}{2!} \frac{d^2 \rho_d(t)}{d t^2} \tau^2+\cdots .
\label{eq3.2}
\end{equation}
As $g\rightarrow0$, only the first term of the expansion contributes to the integral of Eq.(\ref{eq3.1}), since then the effective integration region is confined to the vicinity of $\tau \approx 0$. However, for small coupling, the contributions from the terms with higher powers of $g$ must be taken into account, which correspond to higher-order terms in the expansion (\ref{eq3.2}).

To clarify this point, we substitute the expansion in Eq.(\ref{eq3.2}) into Eq.(\ref{eq3.1}) and explicitly write the evolution equations for the detector's density matrix:
\begin{widetext}
\begin{equation}
\begin{aligned}
 g^{-2}\dot{\rho}_{11}(t)&=  \int_{-\infty}^{\infty} d \tau \mathcal{W}(\tau) e^{-i \omega \tau}-4  \int_0^{\infty} d \tau \operatorname{Re}[\mathcal{W}(\tau)] \cos \omega \tau\bigg(\rho_{11}(t) \underbrace{-\tau\dot{\rho}_{11}(t)+\cdots}_{\text {error }}\bigg) \\
g^{-2}\dot{\rho}_{12}(t)&=-2  \int_0^{\infty} d \tau \operatorname{Re}[\mathcal{W}(\tau)] e^{i \omega \tau}\bigg(\rho_{12}(t) \underbrace{-\tau \dot{\rho}_{12}(t)+\cdots }_{\text {error }}\bigg) 
+2  e^{2 i \omega t} \int_0^{\infty} d \tau \operatorname{Re}[\mathcal{W}(\tau)] e^{-i \omega \tau}\bigg(\rho_{12}^*(t) \underbrace{-\tau \dot{\rho}^*_{12}(t)+\cdots }_{\text {error }}\bigg)
\end{aligned}.
\label{eq3.3}
\end{equation}
\end{widetext}
These equations permit iterative solutions. For instance, when the first-order error part of the first equation in Eqs.(\ref{eq3.3}) contains $\dot{\rho}_{11}(t)$, substituting it directly back into the solution yields its lowest-order contribution of order $g^4$. Likewise, the second-order error term, which includes $\ddot{\rho} (t)$, results in a lowest-order contribution of order $g^6$. This convinces us that if we take into account higher-order error terms, their contributions will correspond to higher powers of $g^2$.

After some calculations, Eqs. (\ref{eq3.3}) become
\begin{widetext}
\begin{equation}
\begin{aligned}
g^{-2}\dot{\rho}_{11}(t)= & - \gamma\omega-2  C_{\Omega} \rho_{11}(t)+2  \frac{d \Delta_{\Omega}}{d \omega} \dot{\rho}_{11}(t), \\
g^{-2}\dot{\rho}_{12}(t)= & -\left(C_{\Omega}+i\Delta_{\Omega}\right) \rho_{12}(t)+e^{2 i \omega t} \left(C_{\Omega}-i \Delta_{\Omega}\right) \rho_{12}^*(t) 
 +\left(\frac{d \Delta_{\Omega}}{d \omega}-i \frac{d C_{\Omega}}{d \omega}\right) \dot{\rho}_{12}(t)
 - e^{2 i \omega t}\left(\frac{d \Delta_{\Omega}}{d \omega}+i \frac{d C_{\Omega}}{d \omega}\right) \dot{\rho}^*_{12}(t),
\end{aligned}   
\label{eq3.4}
\end{equation}
where for brevity, we introduce the notations,
\begin{equation}
\begin{aligned}
& C_{\Omega}:=2[\gamma(\omega)+\gamma(-\omega)]=2 \int_0^{\infty} d \tau \operatorname{Re}[\mathcal{W}(\tau)] \cos \omega \tau, \\
& \Delta_{\Omega}:=S(\omega)-S(-\omega)=2 \int_0^{\infty} d \tau \operatorname{Re}[\mathcal{W}(\tau)] \sin \omega \tau.
\end{aligned}
\label{eq3.5}
\end{equation}
By iterating Eqs. (\ref{eq3.4}) and truncating to order $g^4$, we obtain
\begin{equation}
\begin{aligned}
& g^{-2}\dot{\rho}_{11}(t)=\left[ \gamma(-\omega)-2 C_{\Omega} \rho_{11}(t)\right]\left(1+2 g^2 \frac{d \Delta_{\Omega}}{d w}\right) ,\\
&g^{-2} \dot{\rho}_{12}(t)=\left[e^{2 i \omega t} \left(C_{\Omega}-i \Delta_{\Omega}\right) \rho_{12}^*-\left(C_{\Omega}+i\Delta_{\Omega}\right) \rho_{12}\right]\left(1-2 i g^2 \frac{d C_\Omega}{d w}\right).
\end{aligned}
\label{eq3.6}
\end{equation}
\end{widetext}

To determine the valid range of the Markov approximation, we require that the influence of the first-order error term in Eqs.(\ref{eq3.3}) is sufficiently smaller than the other ``Markov" terms in the equation. Then from Eqs.(\ref{eq3.6}), we obtain the following constraints:
\begin{equation}
g^2\left|\frac{d \Delta_{\Omega}}{d \omega}\right| \ll 1, \quad\quad g^2\left|\frac{d C_{\Omega}}{d \omega}\right| \ll 1.
\label{eq3.7}
\end{equation}

This is one of the key results of our paper. Recall that we are interested in the late-time limit $g^2 t \sim \mathcal{O}(1)$ with small $g$, the reliability conditions Eqs.(\ref{eq3.7}) indicates that the Markov approximation becomes accurate after the timescale
\be
t_{\text{Markov}}\gg \left|\frac{d \Delta_{\Omega}}{d \omega}\right|, \quad\quad t_{\text{Markov}}\gg \left|\frac{d C_{\Omega}}{d \omega}\right| .
\ee
One can observe that the valid range given by Eqs. (\ref{eq3.7}) always holds true for sufficiently weak coupling, which is consistent with Davies' theorems. Moreover, Eqs. (\ref{eq3.7}) can be viewed as an extension of the analysis of \cite{sec1-13,sec1-14} to the detector-field model, as it restricts physical parameters to guarantee the Markov approximation holds.

\subsection{Applicability conditions for the RWA}

Under the restrictions (\ref{eq3.7}), Eqs.(\ref{eq3.6}) become
\begin{equation}
\begin{aligned}
g^{-2}\dot{\rho}_{11}(t)&=\gamma(-\omega)-2 C_{\Omega} \rho_{11}(t), \\
g^{-2}\dot{\rho}_{12}(t)&=e^{2 i \omega t} \left(C_{\Omega}-i \Delta_{\Omega}\right) \rho_{12}^*(t)-\left(C_{\Omega}+i\Delta_{\Omega}\right) \rho_{12}(t),
\end{aligned}
\label{eq3.8}
\end{equation}
which is a specific form of the Redfield equation, as mentioned earlier, cannot preserve the \textit{CP} property of the detector's quantum dynamics. 

To demonstrate this, we can recast Eqs.(\ref{eq3.8}) into the GKSL form,
\begin{equation}
\frac{d \rho(t)}{d t}=\sum_{j, k=1}^3 c_{j k}\left(\boldsymbol{F}_j \boldsymbol{\rho}(t) \boldsymbol{F}_k^{\dagger}-\frac{1}{2}\left\{\boldsymbol{F}_k^{\dagger} \boldsymbol{F}_j, \boldsymbol{\rho}(t)\right\}\right),
\label{eq3.9}
\end{equation}
with $\boldsymbol{F}_j=\frac{1}{2} \boldsymbol{\sigma}_j$ given by Pauli matrices. The Kossakowski matrix $\mathfrak{c}=\left[c_{j k}\right]$ can be recognized explicitly as
\begin{equation}
\mathfrak{c}=2g^{2}\left[\begin{array}{ccc}
2  {C}_{\Omega} & \Delta_{\Omega}-i \left(\gamma(-\omega)-C_{\Omega}\right) & 0 \\
\Delta_{\Omega}+i \left(\gamma(-\omega)-C_{\Omega}\right) & 0 & 0 \\
0 & 0 & 0
\end{array}\right].
\label{eq3.10}
\end{equation}

If Eq.(\ref{eq3.9}) is the correct QMME, like (\ref{eq2.21}), the Kossakowski matrix must be Hermitian and positive \cite{sec1-13,sec1-14}. However, this cannot be true since among the three eigenvalues of the matrix (\ref{eq3.10}),
\begin{equation}
\begin{aligned}
& \lambda_1=0, \\
& \lambda_2=2 g^2\left(C_{\Omega}+\sqrt{C_{\Omega}^2+\left(\gamma(-w)-C_{\Omega}\right)^2+\Delta_{\Omega}^2}\right), \\
& \lambda_3=2 g^2\left(C_{\Omega}-\sqrt{C_{\Omega}^2+\left(\gamma(-w)-C_{\Omega}\right)^2+\Delta_{\Omega}^2}\right),
\end{aligned}
\label{eq3.11}
\end{equation}
$\lambda_3$ is always negative. This proves that Eqs. (\ref{eq3.8}) cannot preserve the \textit{CP} property, i.e., the density matrix can become nonpositive and, thus, physically unreliable. In other words, the RWA is still necessary to convert Eqs. (\ref{eq3.8}) to a correct QMME.\footnote{While our analysis aligns with Davies' theorems, yielding $g$-dependent reliable conditions for Markov and RWA approximations. It is worth noting that series studies \cite{sec1-33,sec1-34,sec1-35,sec1-36} offer $g$-independent conditions from an effective field theory (EFT) perspective, where negative $\lambda_3$ approaches zero, thereby resolving the non-CP problem without employment RWA.}

We aim to derive RWA's applicability conditions by estimating its arising error in a manner analogous to the Markov approximation. However, we must note that essentially, RWA ignores the oscillatory term contribution in the Redfield equation after time integration. Therefore, what we truly need to assess is the difference between the solutions of Eqs.(\ref{eq3.8}) and the QMME (\ref{eq2.21}), which is RWA applied.

Performing the expansion (\ref{eq3.2}), Eq.(\ref{eq2.21}) becomes
\begin{equation}
\begin{aligned}
& \dot{\rho}_{11}(t)=g^2( \gamma(-\omega)-2 C_{\Omega} \rho_{11}(t)), \\
& \dot{\rho}_{12}(t)=-g^2\left(C_{\Omega}+i\Delta_{\Omega}\right) \rho_{12}(t).
\end{aligned}
\label{eq3.12}
\end{equation}
One can observe that after applying RWA, the only difference to the Redfield equation (\ref{eq3.8}) is that the oscillatory term $e^{2 i \omega t} g^2\left(C_{\Omega}-i \Delta_{\Omega}\right) \rho_{12}^*(t)$ has disappeared.

We are now positioned to compare the solutions of Eqs. (\ref{eq3.12}) and Eqs.(\ref{eq3.8}). First, we note that the second equation of (\ref{eq3.12}) is fairly straightforward to solve, yielding,
\begin{equation}
\rho_{12}(t)=e^{-g^2\left(C_{\Omega}+i \Delta_{\Omega}\right) t}  \rho_{12}(0) .
\label{eq3.17}
\end{equation}
Secondly, to resolve the $\rho_{12}$ in Eqs.(\ref{eq3.8}), it is more convenient to absorb the oscillating factor via $\rho_{12}^{(S)}(t)=e^{-i \omega t} \rho_{12}(t)$ first:
\begin{equation}
\begin{aligned}
\frac{d \rho_{12}^{(S)}(t)}{d t} =& -i \omega \rho_{12}^{(S)}(t)-g^2\left(C_{\Omega}+i\Delta_{\Omega}\right) \rho_{12}^{(S)}(t)\\
&+ g^2\left(C_{\Omega}-i \Delta_{\Omega}\right) \rho_{12}^{(S)*}(t),
\end{aligned}
\label{eq3.13}
\end{equation}
where the superscript $S$ indicates that we are working in the Schr\"odinger picture. To find the solution of Eq.(\ref{eq3.13}) is equivalent to solving the following equation:
\begin{equation}
\frac{d}{d t}
\left(\begin{array}{c}
\rho_{12}^{(S)}(t) \\
\rho_{12}^{{(S)}^*}(t)
\end{array}\right)
=\mathcal{D}
\left(\begin{array}{c}
\rho_{12}^{(S)}(t) \\
\rho_{12}^{{(S)}^*}(t)
\end{array}\right),
\label{eq3.14}
\end{equation}
with
\begin{equation}
\mathcal{D}=\left(\begin{array}{cc}
-g^2 C_{\Omega}-i\left(\omega+g^2 \Delta_{\Omega}\right) & g^2\left(C_{\Omega}-i \Delta_{\Omega}\right) \\
g^2\left(C_{\Omega}+i \Delta_{\Omega}\right) & -g^2 C_{\Omega}+i\left(\omega+g^2 \Delta_{\Omega}\right)
\end{array}\right).
\end{equation}
Analytically, we can resolve Eq.(\ref{eq3.14}) and reverting it back to the interaction picture gives
\begin{widetext}
\begin{equation}
\rho_{12}(t)=e^{\left(i \omega-g^2 C_{\Omega}\right) t}\left[\rho_{12}(0)\left(e^{-i\Sigma t} +i\frac{\Sigma-\omega-g^2\Delta_\Omega}{\Sigma}\sin \Sigma t\right)- i g^2 \rho_{12}^{*}(0) \frac{\left(\Delta_\Omega+i C_\Omega\right)}{\Sigma} \sin \Sigma t\right],
\label{eq3.15}
\end{equation}
\end{widetext}
where
\begin{equation}
\Sigma=\omega \sqrt{1+2 g^2 \frac{\Delta_{\Omega}}{\omega}-g^4\left(\frac{C_{\Omega}}{\omega}\right)^2}.
\label{eq3.16}
\end{equation}

By comparing Eq.(\ref{eq3.17}) and Eq.(\ref{eq3.15}), we aim to determine the conditions under which these two equations are equivalent. First, we observe that the dependency on $\rho_{12}^*(0)$ in Eq.(\ref{eq3.15}) disappears in Eq.(\ref{eq3.17}), which implies
\begin{equation}
g^2\left|\frac{\Delta_{\Omega}}{\Sigma}\right| \ll 1, \quad\quad g^2\left|\frac{C_{\Omega}}{\Sigma}\right| \ll 1,
\label{bujia1.2}
\end{equation}
and leaves Eq.(\ref{eq3.15}) as
\begin{widetext}
\begin{equation}
\begin{aligned}
\rho_{12}(\tau) & \approx e^{i {\left(\omega-\Sigma\right)}t} e^{-C_{\Omega} g^2t} \rho_{12}(0) +e^{i {\omega}t} e^{-C_{\Omega} g^2t}  \rho_{12}(0) \frac{i\left(\Sigma-\omega-g^2 \Delta_\Omega\right)}{\Sigma} \sin {\Sigma t}\\
& \approx  e^{-\left(C_\Omega+i \Delta_\Omega\right) g^2t+i g^2\left( \frac{C_\Omega^2+\Delta_\Omega^2}{2 \omega} \right)g^2t+\mathcal{O}\left(g^6\right)} \rho_{12}(0),\\
& \approx  e^{-\left(C_\Omega+i \Delta_\Omega\right) g^2t}\rho_{12}(0).
\end{aligned}
\label{bujia1.5}
\end{equation}
\end{widetext}
Note that during the derivation, we used the expansion,
\begin{equation}
\Sigma=\omega+g^2 \Delta_{\Omega}-g^4 \frac{C_{\Omega}^2}{2 \omega}-g^4 \frac{\Delta_{\Omega}^2}{2 \omega}+\mathcal{O}\left(g^6\right),
\label{bujia1.6}
\end{equation}
which, in the late-time limit $g^2t\sim\mathcal{O}(1)$, allows us to safely omit the terms with high-order $
\mathcal{O}(g^4)$ in the exponential, and ultimately arrive at the solution Eq. (\ref{eq3.17}).

By expanding Eq.(\ref{bujia1.2}) again using Eq.(\ref{bujia1.6}) and neglecting higher-order terms, we ultimately obtain,
\begin{equation}
g^2\left|\frac{\Delta_{\Omega}}{\omega}\right| \ll 1, \quad\quad g^2\left|\frac{C_{\Omega}}{\omega}\right| \ll 1,
\label{bujia1.3}
\end{equation}
which we refer to as the applicability conditions of the RWA. 

Multiplying both sides by $
\omega$, these inequalities reveal how the RWA requires constraining $
\omega$ by $g^2$,
\begin{equation}
\omega \gg g^2\left| \Delta_{\Omega}\right|, \quad\quad \omega \gg g^2\left| C_{\Omega}\right|.
\label{bujia1.1}
\end{equation}
For fixed $g$, to neglect the oscillatory terms, we need $
\omega$ to be sufficiently large. 

However, when $
\omega$ is fixed, the late-time limit also indicates the timescale at which the master equation becomes accurate,
\begin{equation}
t_{\text{RWA}} \gg\left|\frac{\Delta_{\Omega}}{\omega}\right|, \quad t_{\text{RWA}} \gg\left|\frac{C_{\Omega}}{\omega}\right|
\label{bujia1.0}
\end{equation}
In this sense, even for the detector with a small $
\omega$ is (except when $
\omega = 0$, in which case the interaction is already turned off), as long as the coupling is sufficiently weak or after $t_{\text{RWA}}$, the oscillatory terms characterized by $\omega$ can be effectively recognized as ``rapid oscillatory".

In conclusion, we have established the reliability conditions of the Markov approximation (\ref{eq3.7}) and the applicability conditions of RWA (\ref{bujia1.1}), respectively. These new criteria refine the conventionally adopted descriptions of ``small enough" or ``sufficiently long timescale" (e.g., vaguely $t\gg\left |\alpha^{\prime}-\alpha\right|^{-1}$ and $t\gg\tau_B$). The conditions (\ref{eq3.7}) and (\ref{bujia1.1}) limit the valid range for the QMME. It is interesting to note that the background quantum field affects the reliability of the QMME through $C_\Omega$ and $\Delta_{\Omega}$, which are determined by the field's Wightman function. This suggests that in applications of QMME to quantum gravity \cite{add}, the potential compromise of the UV behavior of quantum fields in its reliability requires serious \textit{a priori} scrutiny.

In specific scenarios, such as the accelerating detector discussed in the next section, the coupling constant is a small fixed value. This means that one can apply conditions (\ref{eq3.7}) and (\ref{bujia1.1}) to narrow the range of permissible physical parameters under which the master equation method can effectively predict the open dynamics undergone the Unruh effect.

\section{An illustrated example: the Unruh effect of an accelerating detector}
\label{4}
Considering a UDW detector undergoing uniform acceleration along the $x$-axis in Minkowski spacetime, its trajectory is:
\begin{equation}
x^\mu(t)=\left[\frac{1}{a} \sinh (a t), \frac{1}{a} \cosh (a t), 0, 0\right],
\label{eq4.1}
\end{equation}
where $a$ represents the uniform acceleration of the detector and $t$ denotes the detector's proper time along the trajectory. 

We examine the interaction between the detector and a scalar field. The total Hamiltonian takes the same form of Eq. (\ref{eq2.1}),
\begin{equation}
H=H_d+H_\phi+g H_{\text{int}}.
\end{equation}
Here, the detector is modeled as a two-level atom like $H_d=\frac{1}{2} \omega \sigma_3$, and the free Hamiltonian of the scalar field is given by
\begin{equation}
\begin{aligned}
H =\frac{1}{2} \int\left[\left(\partial_0 \phi(x)\right)^2+\nabla \phi(x) \cdot \nabla \phi(x)+m^2 \phi(x)^2\right] \mathrm{d}^3 x,
\end{aligned}
\end{equation}
where $m$ is the mass of the field. In terms of annihilation and creation operators $a(k)$ and $a^{\dagger}(k)$, the field operator $\phi(x)$ can be Fourier expanded in the momentum space,
\begin{equation}
\phi(x)=\int \frac{\mathrm{d}^3 k}{(2 \pi)^3 2 \omega_k}\left[a(k) \mathrm{e}^{-\mathrm{i} k x}+a^{\dagger}(k) \mathrm{e}^{\mathrm{i} k x}\right],
\end{equation}
with $\omega_k=\left(\mathbf{k}^2+m^2\right)^{1 / 2}$. The interaction Hamiltonian $H_{\text {int}}$ describes the coupling between the detector and the field, taking the form $\mathfrak{m}(0) \otimes \phi[x(t)]$, as we have stated in Sec. \ref{2}, where $\mathfrak{m}(0)$ is the monopole momentum operator of the detector, generally represented as ($\sigma_{+}+\sigma_{-}$). We establish the initial state of the detector and field as
\begin{equation}
|M\rangle\langle M| \otimes \rho_d(0),
\label{eq4.2}
\end{equation}
where $|M\rangle$ corresponds to the Minkowski vacuum.

The Wightman function of the massive field can be calculated as \cite{sec1-4-1,sec4-0-1}
\begin{equation}
\mathcal{W}(t)=\frac{a m}{8 i \pi^2} \frac{K_1\left(\frac{2 m i}{a}\left[\sinh (\frac{a t}{2})-i \frac{a \varepsilon}{2}\right]\right)}{\sinh (\frac{a t}{2})-i \frac{a \varepsilon}{2}} ,
\label{eq4.3}
\end{equation}
where $m$ indicates the mass of the scalar field, $K_\nu(x)$ is the modified Bessel function of the second kind, and $\varepsilon$ is the regulator. 

In the massless limit, we have
\begin{equation}
\mathcal{W}_0(t):=\lim _{m \rightarrow 0} \mathcal{W}(t)=-\frac{a^2}{16 \pi^2\left[\sinh (a t / 2)-i \frac{a \varepsilon}{2}\right]^2}.
\label{eq4.4}
\end{equation}

The Wightman functions Eq. (\ref{eq4.3}) and Eq. (\ref{eq4.4}) exhibit a periodic nature with respect to imaginary time, satisfying
\begin{equation}
\mathcal{W}(t-i \beta)=\mathcal{W}(-t),
\label{eq4.5}
\end{equation}
known as the Kubo-Martin-Schwinger (KMS) condition \cite{sec4-0-2,sec4-0-3}. This hallmark feature ensures that the transition rate of the detector from the ground state will ultimately reach equilibrium, from which an effective Unruh temperature proportional to its acceleration can be recognized from the detector's Planckian spectrum,
\begin{equation}
T:=\frac{1}{\beta}=\frac{a}{2 \pi}.
\label{eq4.6}
\end{equation}
That is, the accelerating UDW detector in Minkowskian spacetime perceives thermal radiation at Unruh temperature (\ref{eq4.6}) and ultimately thermalizes to an equilibrium Gibbs state. This encapsulates the essence of the initially proposed version of the Unruh effect \cite{sec1-2}.

Reaching a unique thermalization end determined solely by the detector's acceleration 
\cite{sec1-4-1} does not imply that the detector follows the same thermalization path within its Hilbert space. Instead, the detector's thermalization path varies based on the background quantum field being studied and the specific setup of the detector (such as its energy spacing or initial state preparation) \cite{sec4-0-4,sec4-0-5}. To fully explore the quantum nature of the Unruh effect, the open dynamics of the UDW detector have first been examined using the QMME (\ref{eq3.12}) in \cite{sec1-17}. This open quantum system approach then allows one to directly ascertain if and how a UDW detector undergoing the Unruh effect approaches a Gibbs state, rather than stopping at the Planckian transition rates, which cannot capture the decoherence of the detector's density matrix, and thus serve only as necessary but not sufficient conditions for detector thermalization.

To see this, we first solve the quantum Markovian master equation (\ref{eq3.12}),
\begin{equation}
\rho_{11}(t)=e^{-2 g^2 C_{\Omega} t}\left(\rho_{11}(0)-\frac{\gamma(-w)}{2 C_{\Omega}}\right)+\frac{\gamma(-w)}{2 C_{\Omega}},
\label{eq4.8}
\end{equation}
and
\begin{equation}
\rho_{12}(t)=\rho_{12}(0) e^{-g^2\left(C_\Omega+i \Delta_{\Omega}\right) t}.
\label{eq4.9}
\end{equation}
By Eq.(\ref{eq2.24}) and Eq.(\ref{eq3.5}), the definitions of $\gamma(-w)$ and $C_{\Omega}$, one can verify that when the Wightman function satisfies the KMS condition (\ref{eq4.5}), the solution of (\ref{eq4.8}) can be written as
\begin{equation}
\rho_{11}(t)=e^{-2 g^2 C_\Omega t}\left(\rho_{11}(0)-\frac{1}{e^{\beta \omega}+1}\right)+\frac{1}{e^{\beta \omega}+1}.
\label{eq4.10}
\end{equation}

As time passes, the detector undergoes thermalization due to Unruh radiation. After a sufficiently long duration, by combining Eq.(\ref{eq4.9}) and Eq.(\ref{eq4.10}), we obtain the asymptotic state of the detector, which is a Gibbs state,
\begin{equation}
\lim _{t \rightarrow \infty} \rho_d(t)=\left[\begin{array}{cc}
\frac{1}{e^{\beta \omega}+1} & 0 \\
0 & \frac{1}{e^{-\beta \omega}+1}
\end{array}\right]=\frac{e^{-\beta H_d}}{\operatorname{Tr}\left[e^{-\beta H_d}\right]}=\rho_{\infty}.
\label{eq4.11}
\end{equation}
It is evident that the asymptotic state relies exclusively on the Unruh temperature or the detector's acceleration, demonstrating the prediction of the Unruh effect in the ``conventional field approach". 

Now, regarding the accelerating UDW detector model, we return to the theme of this paper; while the open dynamics encodes the thermalization process experienced during the Unruh effect, under what conditions is the dynamics derived from the QMME valid? Utilizing previously established conditions (\ref{eq3.7}) and (\ref{bujia1.3}), we can address this issue for both massless and massive backgrounds, respectively.

\subsection{The massless case}
To examine the condition (\ref{eq3.7}) and condition (\ref{bujia1.3}) for accelerating UDW detector, we first need to know the value of $C_{\Omega}$ and $ \Delta_{\Omega}$ which are given as \cite{sec1-33}
\begin{equation}
C_{\Omega}=\lim _{\varepsilon \rightarrow 0^{+}} 2 \int_0^{\infty} d \tau \operatorname{Re}[\mathcal{W}(\tau)] \cos \omega \tau=\frac{\omega}{4 \pi} \operatorname{coth}\left(\frac{\pi \omega}{a}\right),
\label{eq4.12}
\end{equation}
and
\begin{equation}
\Delta_{\Omega}=2 \int_0^{\infty} d \tau \operatorname{Re}[\mathcal{W}(\tau)] \sin \omega \tau \approx  \frac{\omega}{2 \pi}\log \left(e^\gamma \omega \varepsilon\right),
\label{eq4.13}
\end{equation}
where $\gamma$ is the Euler-Mascheroni constant. 

It is important to note that $\Delta_{\Omega}$ diverges once the regulator $\varepsilon$ be taken to the rigorous limit as $\varepsilon \rightarrow 0$ in Eq.(\ref{eq4.13}). This indicates that the detector is sensitive to the ultraviolet behavior of the quantum field, which is not surprising since the UDW detector is an effective model and would lose its validity at higher energy scales. Nevertheless, we can admit an alternative interpretation of the regulator $\varepsilon$ as the spatial profile of the detector \cite{sec4-1,sec4-2,sec4-3,sec4-4}. A detector with a finite spatial profile would not be sensitive to the short-distance behavior of the quantum field, thus avoiding divergences.\footnote{However, we must note that such a UV regulator, implemented through the finite detector spatial profile, cannot function at arbitrary scales. In particular, the UV divergences will inevitably reappear when taking the pointlike limit for the detector. } In the following, we use $\varepsilon$ to represent the size of the UDW detector.

Substituting Eq.(\ref{eq4.13}) back into Eqs.(\ref{eq3.7}), we obtain the first reliability condition for the Markov approximation:
\begin{equation}
g^2\left|\frac{d \Delta_{\Omega}}{d \omega}\right| \approx \frac{g^2}{2 \pi}\left|\log \left(e^\gamma \omega \varepsilon\right)\right| \ll 1.
\label{eq4.15}
\end{equation}
We observe that even for weak coupling with a small $g^2$, the combination $\omega \varepsilon$ of the detector energy spacing and its size cannot be too small; otherwise, it breaks the constraint (\ref{eq4.15}), causing the arising error under the Markov approximation to diverge.

\begin{figure}[t] 
\begin{center}
    \includegraphics[width=.47\textwidth]{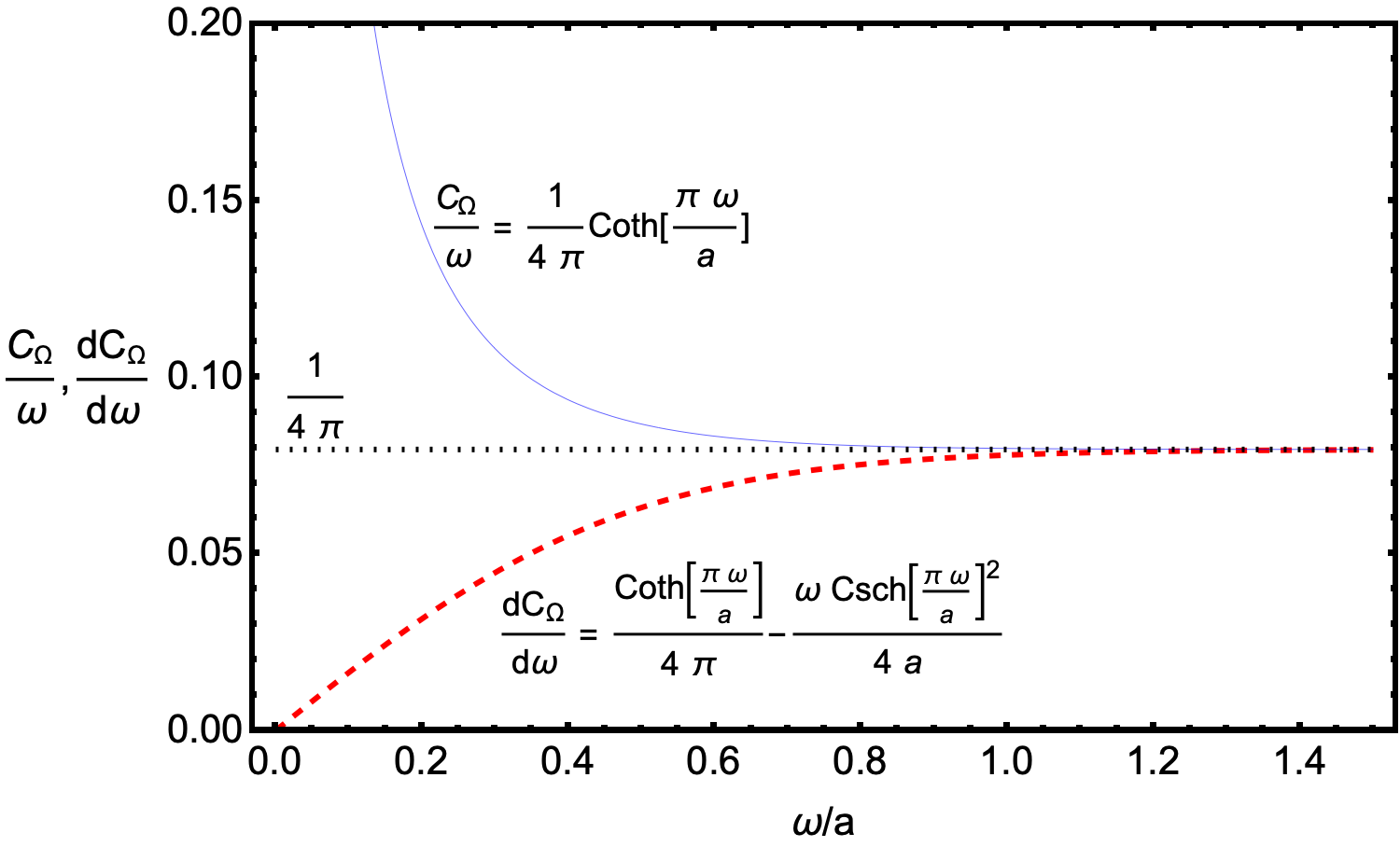}
    \caption{The dependency of the functions ${d C_{\Omega}}/{d \omega}$ (red dashed line) and ${C_{\Omega}}/{\omega}$ (blue solid line) on $\omega / a$ in the massless case. Both functions approach $1 / 4 \pi $ at the limit $\omega \gg a$. In the region of $\omega \ll a$, ${d C_{\Omega}}/{d \omega}$ approaches zero, while ${C_{\Omega}}/{\omega}$ diverges. This contradiction can be reconciled with a small coupling constant to uphold the Markov approximation and RWA simultaneously.}
        \label{fig1}
    \end{center}
\end{figure}

By substituting Eq.(\ref{eq4.12}) into Eqs.(\ref{eq3.7}), we obtain the second reliability condition for Markov approximation as:
\begin{equation}
g^2\left|\frac{d C_{\Omega}}{d \omega}\right|=g^2\left[\frac{1}{4 \pi} \operatorname{coth}\left(\frac{\pi \omega}{a}\right)-\frac{\omega}{4 a} \operatorname{csch}\left(\frac{\pi \omega}{a}\right)^2\right] \ll 1.
\label{eq4.14}
\end{equation}
From the left side of the inequality, it is clear that ${d C_{\Omega}}/{d \omega}$ depends solely on the ratio $\omega / a$, which is illustrated by the red dashed line in Fig. \ref{fig1}. In the region where $\omega \ll a$, the function ${d C_{\Omega}}/{d \omega}$ approaches zero, indicating that the validity of the Markov approximation is firmly assured in this region. On the other hand, for $\omega \gg a$, the magnitude of ${d C_{\Omega}}/{d \omega}$ is bounded to $1 / 4 \pi $, which implies that even when the detector's energy spacing significantly exceeds its acceleration, there is no need to worry about the potential breakdown of the Markov approximation. 

We next proceed toward the applicability of the RWA (\ref{bujia1.3}) which can be given explicitly by plugging Eq.(\ref{eq4.12}) and Eq.(\ref{eq4.13}) as
\begin{equation}
\begin{aligned}
 g^2\left|\frac{\Delta_{\Omega}}{\omega}\right| &\approx \frac{g^2}{2 \pi}\left|\log \left(e^\gamma \omega \varepsilon\right)\right| \ll 1,\\
 g^2\left|\frac{C_{\Omega}}{\omega}\right|&=\frac{g^2}{4 \pi} \operatorname{coth}\left(\frac{\pi \omega}{a}\right) \ll 1.
\label{eq4.16}
\end{aligned}
\end{equation}

The first inequality is the same as Eq.(\ref{eq4.15}), so it does not offer any new information. The left side of the second inequality is illustrated by the blue solid line in Fig.\ref{fig1}. We see ${C_{\Omega}}/{\omega}$ diverges when $\omega$ is much smaller than $a$. This occurs because, in the small $\omega / a$ region, the high-frequency oscillatory terms neglected by the RWA appear to oscillate at lower frequencies compared to the detector's typical evolution timescale $\mathcal{O}\left(1 / g^2\right)$, leading to the breakdown of the RWA. At first glance, this behavior of the function ${C_{\Omega}}/{\omega}$ seems somewhat in conflict with the reliability condition (\ref{eq4.14}) for the Markov approximation. However, noting from (\ref{eq4.15}) and (\ref{eq4.16}) that both the values of ${d C_{\Omega}}/{d \omega}$ and ${C_{\Omega}}/{\omega}$ are suppressed by $g^2$ indicates that one can adjust a sufficiently small coupling constant $g$ to extend the detector's typical evolution timescale and cause the oscillatory terms to behave like high-frequency again. In this way, the RWA is secured, and the contradiction of upholding the Markov approximation and RWA simultaneously is reconciled.

In summary, the open dynamics of an accelerating UDW detector interacting with a massless background can be reliably described by the QMME, with additional constraints on small detector size and low-energy gap regions. Furthermore, a sufficiently small coupling constant can effectively suppress the magnitude of the error terms.

\subsection{The massive case}

For a massive quantum field, the Wightman function takes an intricate form, as shown in (\ref{eq4.3}). By substituting this into the definition (\ref{eq3.5}), we can straightforwardly give $C_\Omega$ as
\begin{widetext}
\begin{equation}
\begin{aligned}
C_{\Omega}  =\lim _{\varepsilon \rightarrow 0^{+}} 2 \int_0^{\infty} d \tau \operatorname{Re}[\mathcal{W}(\tau)] \cos \omega \tau 
 =\frac{m^2 \cosh \left({\pi \omega}/{a}\right)}{4 \pi^2 a}\left\{K_{\frac{i \omega}{a}-1}\left({m}/{a}\right) K_{\frac{i \omega}{a}+1}\left({m}/{a}\right)
-\left[K_{\frac{i w}{a}}\left({m}/{a}\right)^2\right]\right\}.
\end{aligned}
\label{eq4.17}
\end{equation}

The calculation of $\Delta_\Omega$ is more subtle. Note that in the limit $t\rightarrow0$, the massive Wightman function exhibits the same asymptotic form as the massless case,
\begin{equation}
\mathcal{W}_{\text{asymp}}(t):=\lim_{t\rightarrow 0} \mathcal{W}(t)\sim \frac{1}{(t-i \varepsilon)^2}.
\label{eq4.19}
\end{equation}
We can divide the integral in (\ref{eq3.5}) into the divergent and convergent parts of $\mathcal{W}(\tau)$ \cite{sec1-33}, respectively,
\begin{equation}
\begin{aligned}
\Delta_{\Omega} & =2 \lim _{\varepsilon \rightarrow 0^{+}} \int_0^{\infty} d \tau \operatorname{Re}[\mathcal{W}(\tau)] \sin \omega \tau \\
& =2 \int_0^{\infty} d \tau \operatorname{Re}\left[\mathcal{W}_{\text{asymp}}(\tau)\right] \sin \omega \tau+2 \lim _{\varepsilon \rightarrow 0^{+}} \int_0^{\infty} d \tau \operatorname{Re}\left[\mathcal{W}(\tau)-\mathcal{W}_{\text{asymp}}(\tau)\right] \sin \omega \tau \\
& =\frac{\omega}{2 \pi} \log \left(e^\gamma \omega \varepsilon\right)+\mathcal{O}\left(\omega^2 \varepsilon^2\right)+\text {convergent term},  
\end{aligned}
\label{eq4.20}
\end{equation}
\end{widetext}
where the second term in the second step no longer has the singularity, allowing us to safely take $\varepsilon \rightarrow 0^{+}$ and refer to its integral as the ``convergent term". From Eq. (\ref{eq4.20}), we can see that the value of $\Delta_\Omega$ primarily depends on the specific form of $\mathcal{W}_{\text {asymp}}(t)$, which remains unchanged in both the massive and massless cases. Consequently, we have the function $\Delta_\Omega$ for massive case,
\begin{equation}
\Delta_{\Omega}=2 \int_0^{\infty} d \tau \operatorname{Re}[\mathcal{W}(\tau)] \sin \omega \tau \approx  \frac{\omega}{2 \pi}\log \left(e^\gamma \omega \varepsilon\right),
\label{eq4.18}
\end{equation}
which is the same as the massless case (\ref{eq4.13}).

We are now ready to explore the reliability conditions for the Markov approximation (\ref{eq3.7}) and the applicability of the RWA (\ref{bujia1.3}), which are determined as
\begin{widetext}
\begin{equation}
\begin{aligned}
 \frac{d C_{\Omega}}{d \omega}=&\frac{m^2 \sinh \left({\pi \omega}/{a}\right)}{4 \pi a^2}\left\{K_{\frac{i \omega}{a}-1}\left({m}/{a}\right) K_{\frac{i \omega}{a}+1}\left({m}/{a}\right)-\left[K_{\frac{i \omega}{a}}\left({m}/{a}\right)\right]^2\right\} \\
 +&\frac{im^2\operatorname{coth}\left({\pi \omega}/{a}\right) }{4 \pi a^2} \left\{K_{\frac{i \omega}{a}+1}\left({m}/{a}\right) K_{\frac{i \omega}{a}-1}^{(1,0)}\left({m}/{a}\right)+K_{\frac{i \omega}{a}-1}\left({m}/{a}\right) K_{\frac{i \omega}{a}+1}^{(1,0)}\left({m}/{a}\right)-2 K_{\frac{i \omega}{a}}\left({m}/{a}\right) K_{\frac{i \omega}{a}}^{(1,0)}\left({m}/{a}\right)\right\},
\end{aligned}
\label{eq4.21}
\end{equation}
and
\begin{equation}
\frac{C_{\Omega}}{\omega} =\frac{m^2\cosh \left({\pi \omega}/{a}\right)}{4 \pi^2 \omega a} \left\{K_{\frac{i \omega}{a}-1}\left({m}/{a}\right) K_{\frac{i \omega}{a}+1}\left({m}/{a}\right)-\left[K_{\frac{i w}{a}}\left({m}/{a}\right)^2\right]\right.,
\label{eq4.22}
\end{equation}
where an abbreviation has been introduced as:
\be
K_{\nu}^{(1,0)}\left(x\right):=\frac{1}{2} \pi \csc (\nu \pi)\left(\frac{\partial I_{-\nu}(z)}{\partial \nu}-\frac{\partial I_\nu(z)}{\partial \nu}\right)-\pi \cot (\nu \pi) K_\nu(z),
\ee
including a derivative of $I_\nu(z)$, the modified Bessel function of the first kind.
\end{widetext}

\begin{figure*}[htbp]
\begin{center}
\begin{minipage}{\textwidth}
\subfloat[]{\includegraphics[width=.45\textwidth]{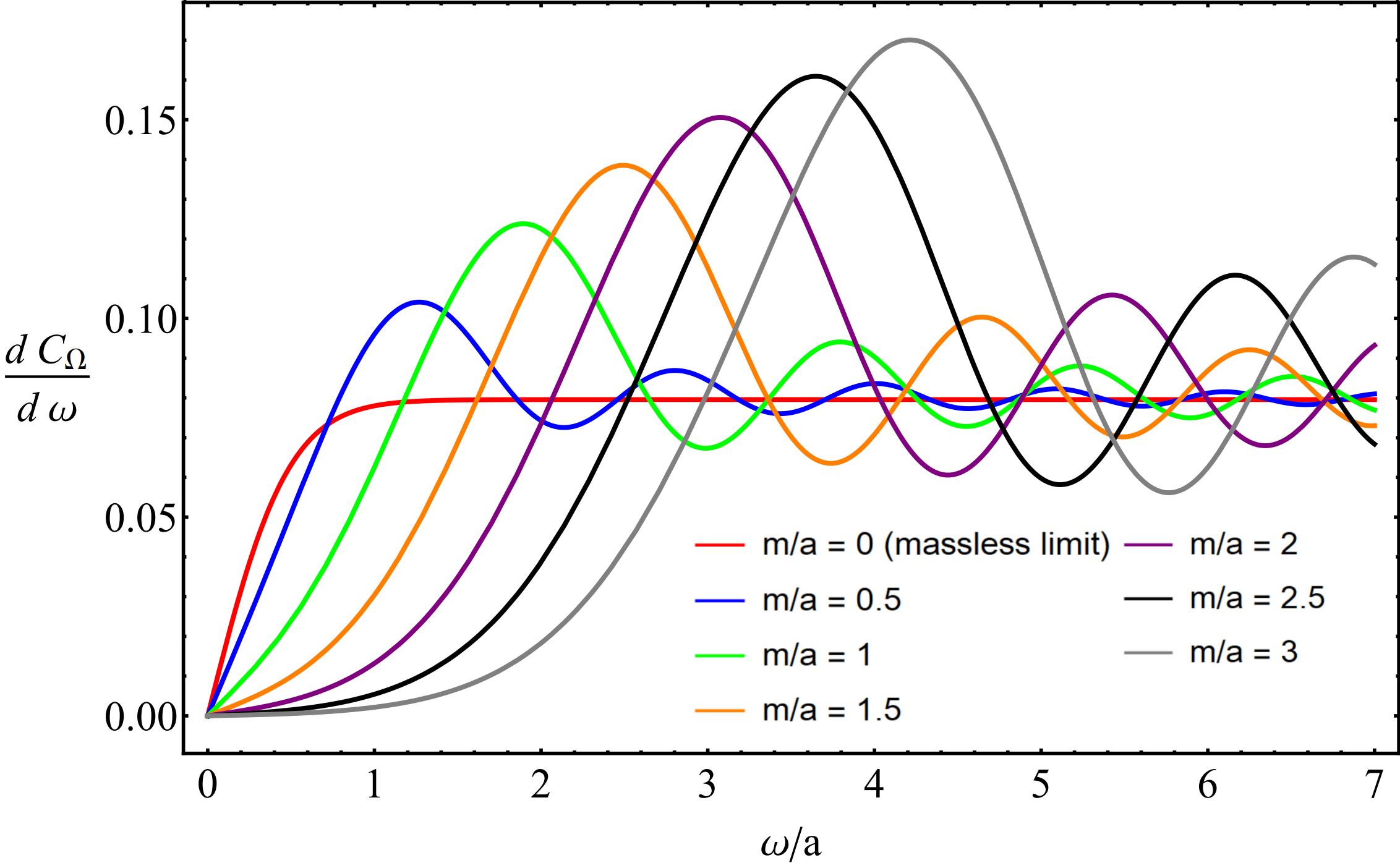}}~~~~~~
\subfloat[]{\includegraphics[width=.45\textwidth]{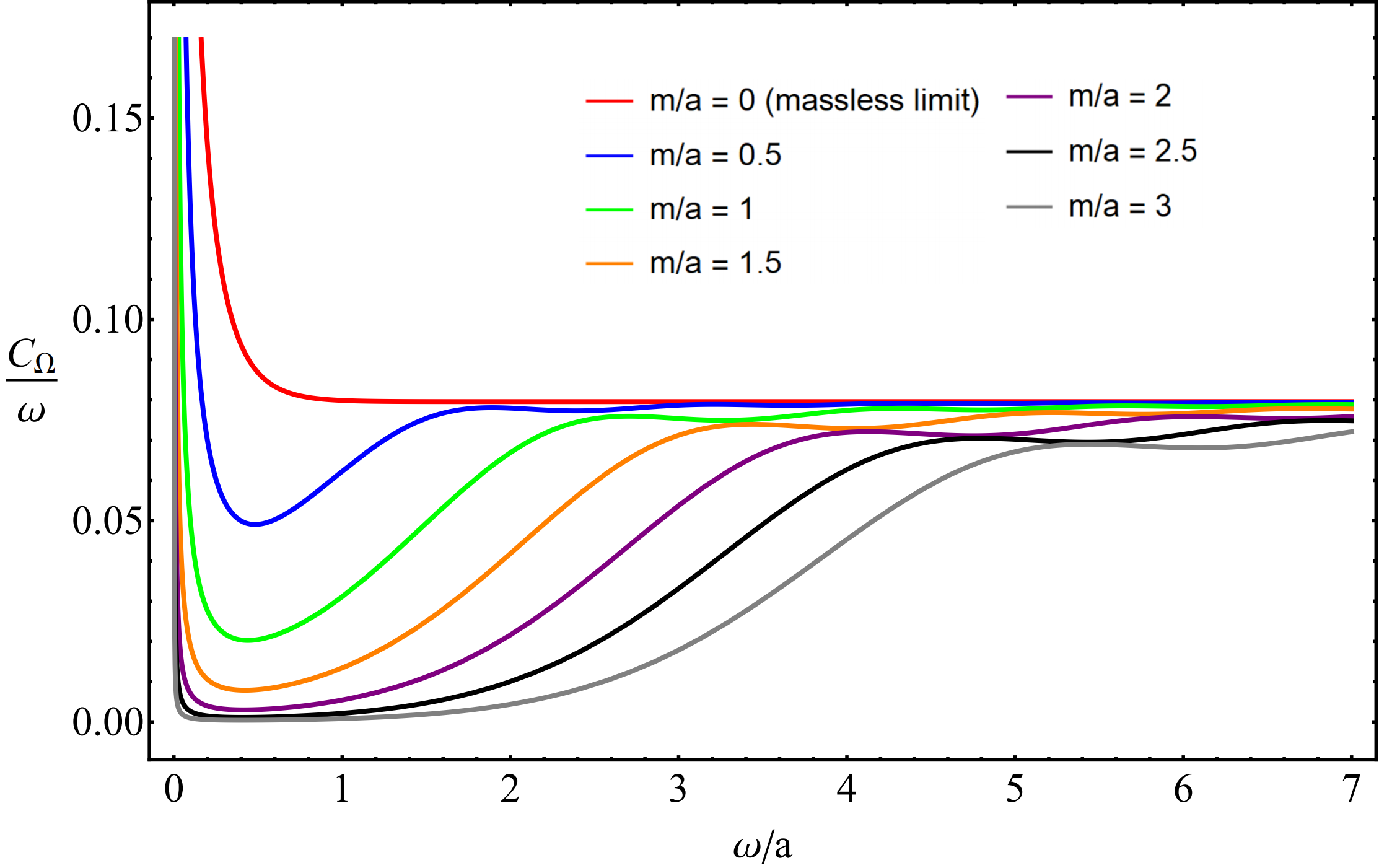}}\\
\subfloat[]{\includegraphics[width=.45\textwidth]{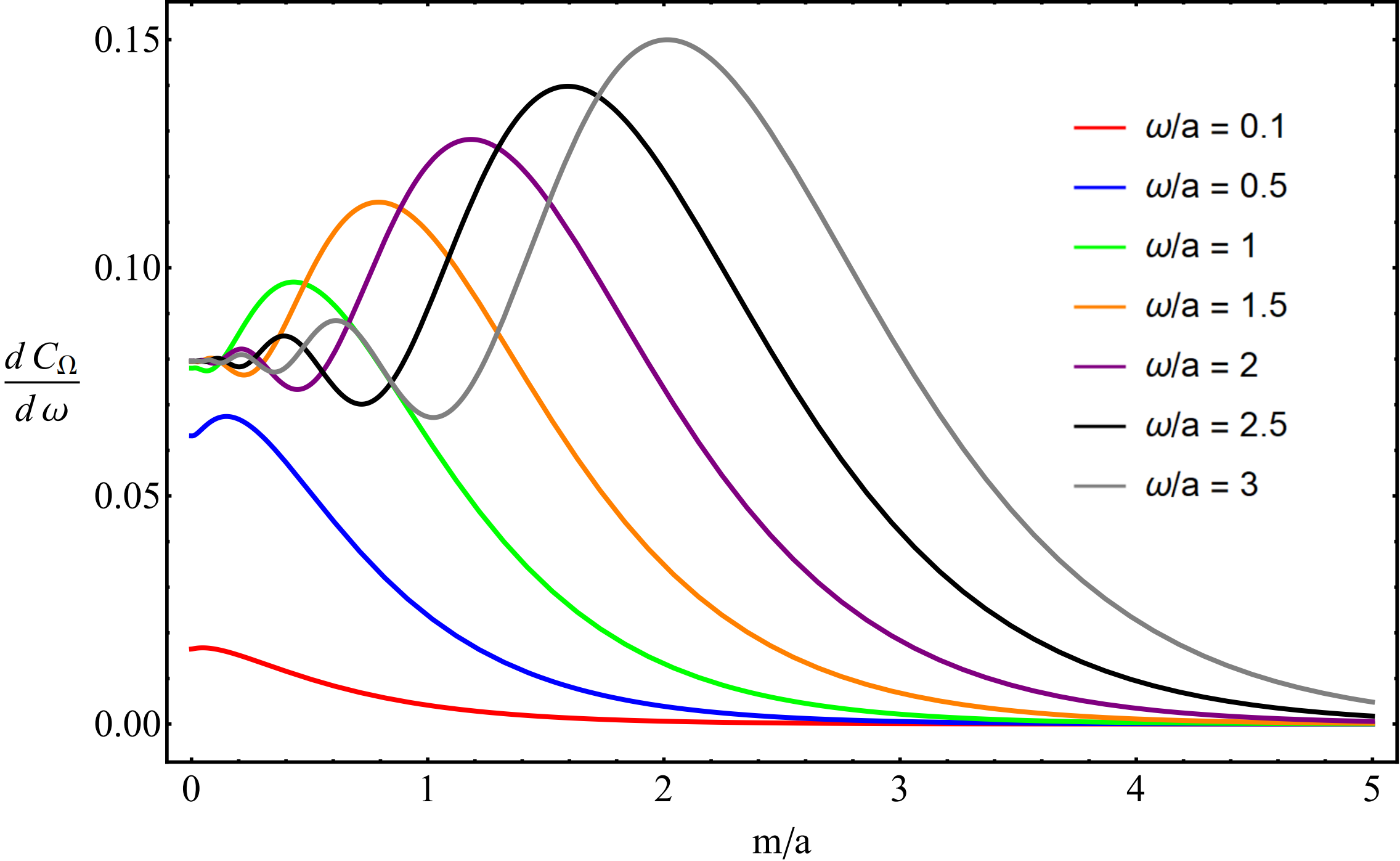}}~~~~~~
\subfloat[]{\includegraphics[width=.45\textwidth]{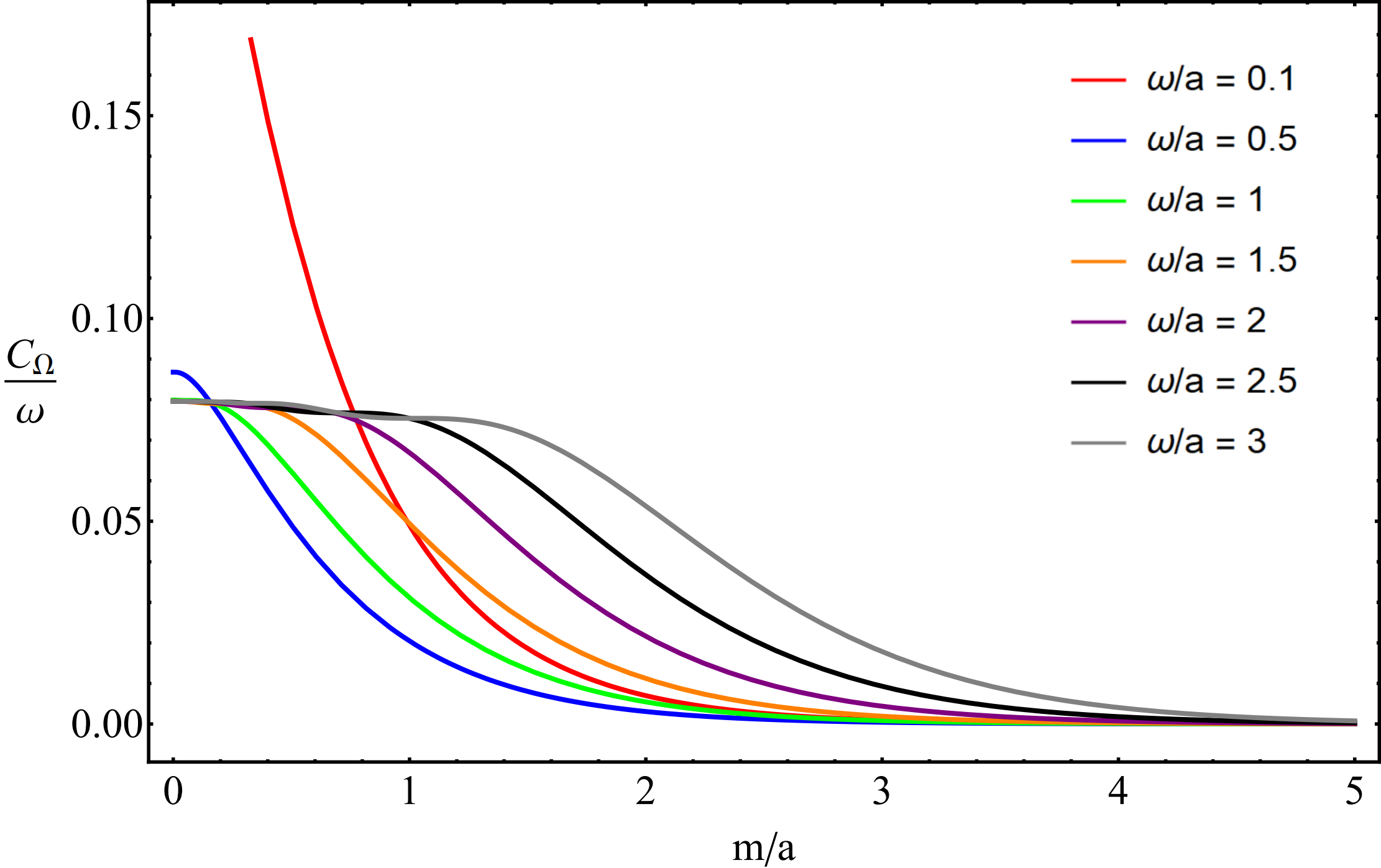}}
\end{minipage}
\caption{(a) For different values of $m / a$, ${d C_{\Omega}}/{d \omega}$ varies as a function of $\omega / a$. All the curves approach zero when $\omega \ll a$, and asymptotically converge to $1/4 \pi$ for sufficiently large $\omega \gg a$. (b) For different values of $m / a$, ${C_{\Omega}}/{\omega}$ varies as a function of $\omega / a$. All the curves diverge at $\omega / a \rightarrow 0$, indicating that without sufficient suppression by a small coupling constant, the RWA may fail in this region. 
(c) For different choices of $\omega / a$, ${d C_{\Omega}}/{d \omega}$ varies as a function of $m / a$. All the curves approach zero for $m/a \gg 1$, indicating the Markov approximation is reliable for heavy quantum field background. (d) For different choices of $\omega / a$, ${C_{\Omega}}/{\omega}$ varies as a function of $m / a$. All the curves approach zero for $m/a \gg 1$, indicating the RWA is applicable for heavy quantum field background.}
\label{fig2}
\end{center}
\end{figure*}

Both ${d C_{\Omega}}/{d \omega}$ and ${C_{\Omega}}/{\omega}$ have two independent dimensionless variables, $\omega/a$ and $m/a$, as illustrated in Fig.\ref{fig2}. From Fig.\ref{fig2}(a) and Fig.\ref{fig2}(c), we observe that the values of ${d C_{\Omega}}{d \omega}$ are bounded. Therefore, similar to the massless case, the Markov approximation is reliable, especially becoming accurate in the regions $m / a \gg 1$ or $\omega / a \ll 1$ (with ${d C_{\Omega}}/{d \omega} \rightarrow 0$).

However, as seen in Fig.\ref{fig2}(b), the function ${C_{\Omega}}/{\omega}$ diverges in the region $\omega / a \ll 1$, indicating a failure of RWA. To maintain the validity of employing the RWA and Markov approximation simultaneously, we can require a sufficiently small coupling constant $g$ to suppress the divergence of errors caused by the RWA. Moreover, from Fig.\ref{fig2}(d), we can see that ${C_{\Omega}}/{\omega}$ approaches zero in the region where $m$ is much larger than $a$, which coincides with the Markov approximation. This suggests that the QMME is more reliable for describing the behavior of a detector interacting with a heavier quantum field.

\section{Conclusion}
\label{5}
We present a method to estimate the validity range of the QMME for the UDW detector in a general context, particularly without requiring any exact solution for the detector's evolution. Our key results include reliability conditions (\ref{eq3.7}) for the Markov approximation and applicability conditions (\ref{bujia1.3}) for the RWA. Both validity conditions are $g$-dependent, indicating that they hold for $g\rightarrow0$, consistent with Davies' theorems. Nevertheless, the specific forms of these conditions depend on the details of the setting of an open system. For the UDW detector model, these could be the spacetime background, the trajectory of the detector, and the type of quantum field being studied. We illustrate this by reexamining the open dynamics of an accelerating UDW detector undergoing the Unruh effect, where the valid conditions narrow the parameter space to ensure the reliability of the prediction of the QMME (\ref{eq2.21}). In particular, we found that the errors caused by the Markov and RWA approximations are bounded in most parameter regions without requiring extremely weak coupling. Only in regions with small detector energy spacing or spatial size do the error terms need extremely weak coupling to suppress them and restore the reliability of the QMME.

The approach outlined in this paper is applicable to a variety of scenarios, particularly those with demands on specifying the range of physical parameters required for experimental verification. For instance, the circular motion open system permits the acceleration process to continue for an indefinite interaction duration. This scenario is particularly compelling 
\cite{sec5-1,sec5-2} for directly detecting the Unruh effect within a finite-size laboratory. Theoretically, to reliably predict the detector's long-time evolution via the QMME, it is important to impose restrictions on the physical parameters, which can be fulfilled in a specific experimental arrangement as well \cite{sec5-3}.

Our analysis did not account for potential errors stemming from the Born approximation. This assumption could prove inadequate, particularly when the ``bath" is insufficiently large to disregard the detector's backreaction (such as when the detector engages with small, nonideal thermal reservoirs). It would be important to extend our method in future to assess the specific nature of the errors caused by the Born approximation, completing our investigation into the reliability of the QMME.

\section*{Acknowledgments}
This work is supported by the National Natural Science Foundation of China (No. 12475061 and No. 12075178), the Shaanxi Fundamental Science Research Project for Mathematics and Physics (No. 23JSY006), and the Innovation Program for Quantum Science and Technology (2021ZD0302400).

\appendix
\section{Upper bound of the second term of the part}
\label{A}
To estimate the magnitude of the second term of the error part in Eq.(\ref{eq3.1}), it suffices to bound it by a convenient norm. For arbitrary operator $A$ in a Hilbert space, and define $|A| \equiv \sqrt{A^{\dagger} A}$, we list two useful types of norms:

1. The trace norm:

\be
\|A\|_1 := \operatorname{Tr}|A|=\sum_i s_i(A),
\label{A1}
\ee
where $s_i(A)$ are the singular values of $A$ (i.e., the eigenvalues of $|A|$ ). If $A=\rho$ is a normalized quantum state (e.g., a density matrix), then $\|\rho\|_1=\operatorname{Tr} \rho=1$.

2. The operator norm:

\be
\|A\|_{\infty} \equiv \max _i s_i(A) .
\label{A2}
\ee
By the definition, one has $\|A\|_{\infty} \leq\|A\|_1$ since the largest singular value is one of the summands in $\|A\|_1$.

For convenience, we denote the second term of the error part by $\Delta$. If we expand the commutator in $\Delta$ and consider its Hermitian conjugate, $\Delta$ consists of four parts, which obey exactly the same bound since they differ from each other only in the operator order, which is irrelevant for taking the norm. Therefore, using the triangle inequality and submultiplicativity of the operator norm $\|\cdot\|_{\infty}$ \cite{sec2-1}, we have
\begin{equation}
\begin{aligned}
\|\Delta\|_{\infty} & \leqslant 4 g^2 \int_t^{\infty} d \tau|\mathcal{W}(\tau)|\|\mathfrak{M}(t)\|_{\infty}\|\mathfrak{M}(t-\tau)\|_{\infty}\left\|\rho_d(t-\tau)\right\|_{\infty} \\
& \leqslant 4 g^2 \int_t^{\infty} d \tau|\mathcal{W}(\tau)|\left\|\rho_d(t-\tau)\right\|_1 \\
& =4 g^2 \int_t^{\infty} d \tau|\mathcal{W}(\tau)|,
\end{aligned}
\label{A3}
\end{equation}
where in the second step, we use $\|\mathfrak{M}(t)\|_{\infty}=\|\mathfrak{M}(t-\tau)\|_{\infty}=1$ and in the third step, we use $\left\|\rho_d(t-\tau)\right\|_1=1$, since $\|\rho_d(t-\tau)\|_1=\operatorname{Tr} [\rho_d(t-\tau)]=1$. Taking the van Hove limit, $\|\Delta\|_{\infty}$ becomes zero as $t \rightarrow \infty$. This implies that $\Delta$ is also zero, since $\|\Delta\|_{\infty}$ is its upper bound. 

Next, we want to calculate the order of $\|\Delta\|_{\infty}$ under the late-time limit $g^2 t\sim\mathcal{O}(1)$, which in our analysis replaces the more restrictive van Hove limit. To this aim, we first transform the real-time $t$ to a rescaled time $\mathcal{T}=g^2 t$, so that $\mathcal{T}$ is $\mathcal{O}(1)$, and the inequality (\ref{A3}) becomes
\begin{equation}
\|\Delta\|_{\infty} \leqslant 4 g^2 \int_{\mathcal{T} / g^2}^{\infty} d \tau|\mathcal{W}(\tau)|.
\label{A4}
\end{equation}
Further calculations require the specific form of $\mathcal{W}(\tau)$, which, for now, we can assume has a power-law dependence over time,
\begin{equation}
|\mathcal{W}(\tau)| \sim \frac{1}{\tau^x},
\label{A5}
\end{equation}
which gives the bound of $\|\Delta\|_{\infty}$ as
\begin{equation}
\|\Delta\|_{\infty} \leqslant 4 g^{2 x} \frac{x-1}{\mathcal{T}^{x-1}}.
\label{A6}
\end{equation}
Recall that the leading order of the first term of the error part in Eq.(\ref{eq3.1}) is $g^4$, therefore, when $x>2$, the leading order of the entire error part will appear only in the first term, which safeguards our analysis of the main text. 

It is worth noting that the derived bound (\ref{A6}), and consequently the applicability condition of the Markov approximation (\ref{eq3.7}), heavily relies on the functional characteristics of $\mathcal{W}(\tau)$, particularly its decay rate being faster than the square law. This is a general condition that encompasses most of the interesting cases in quantum gravity, such as the Hawking-Unruh effect discussed in Sec.\ref{4}, where the Wightman function (\ref{eq4.3}) shows square-law decay in the limits of both $a\rightarrow0$ and $m\rightarrow0$, and exponentially decays in all other instances.

\end{document}